# Phase regeneration of QPSK signals based on Kerr soliton combs

Xinjie Han [1], Yong Geng [1,*], Haocheng Ke[1] and Kun Qiu [1]

[1] Key Lab of Optical Fiber Sensing and Communication Networks, University of Electronic Science and Technology of China, Chengdu 611731, China; hanxinjie_uestc@126.com
* Correspondence: gengyong@uestc.edu.cn

**Abstract:** We demonstrate an all-optical phase regeneration technique based on Kerr soliton combs, which can realize degraded quaternary phase shift keying (QPSK) signal regeneration through phase-sensitive amplification. A Kerr soliton comb is generated at the receiver side of optical communication systems based on a carrier recovery scheme and is used as coherent dual pumps to achieve phase regeneration. Our study will enhance the relay and reception performance of all-optical communication systems.

## 1. Introduction

The rapid evolution of high-speed and high-capacity optical communication networks has resulted in an increased burden on optical network nodes in terms of system power consumption due to the photoelectric and electro-optical conversion processes. To address this concern, all-optical signal processing systems have been designed, which operate within the optical domain to process optical information. This approach capitalizes on the advantages of photonics, including high-speed transmission, broadband capacity, low power consumption, and low cost, to significantly enhance the overall performance of communication networks while concurrently reducing energy consumption and operating costs [1,2]. All-optical phase regeneration, as a subcategory of the all-optical signal processing process, can be utilized at both the relay node and the signal receiving end of an optical communication system. This technique primarily employs the phase-sensitive amplification process, which exploits its sensitivity to phase matching to achieve phase noise compression.

In recent years, there has been significant progress in phase-sensitive amplification (PSA) techniques for telecommunications and high-speed signal processing applications. The two primary focus areas have been low-noise PSA [3,4] and phase regeneration in high-order modulation formats. Achieving phase regeneration in high-order modulation formats requires solving two major issues: the selection of nonlinear materials and the frequency matching problem of signal and pump. The most commonly used nonlinear materials for PSA are silicon-based waveguides, highly non-linear fibers (HNLF), and periodically poled $LiNbO_3$ (PPLN) [5-9]. In recent years, there has been increasing interest in the utilization of silicon-based waveguides in photonic and electronic integrated circuits due to their high nonlinear coefficient. At telecommunication wavelengths, however, two-photon absorption (TPA) strongly restricts the pump power that can be used as it leads to the accumulation of free carriers (FC), resulting in high loss[10-12]. To address this issue, a frequently employed approach involves integrating other non-linear materials to facilitate the PSA process [9]. On the other hand, although HNLF can achieve phase regeneration without the help of other high nonlinear materials [13], it is common to encounter interference from stimulated Brillouin scatters (SBS) which restricts the amount of input power that can be launched into them [7,8]. This interference has two manifestations: (1) if the pump power required by the phase-sensitive regeneration process is much higher than the SBS threshold, then the SBS process will hinder the production of the phase-sensitive amplification process, and (2) if the pump power required for the phase-sensitive regeneration process is near the SBS threshold, then the SBS process will introduce a large amplitude noise in the phase-sensitive regeneration. A common solution is to spool the fiber with a linear strain gradient during the preparation of HNLF [14], thereby increasing the SBS threshold and achieving phase-sensitive regeneration. Our proposed approach aims to promote the frequency coherence between signal and the pump by Kerr soliton combs, which enables a phase-sensitive regeneration process in HNLF at a lower pump input power, avoiding interference from SBS.

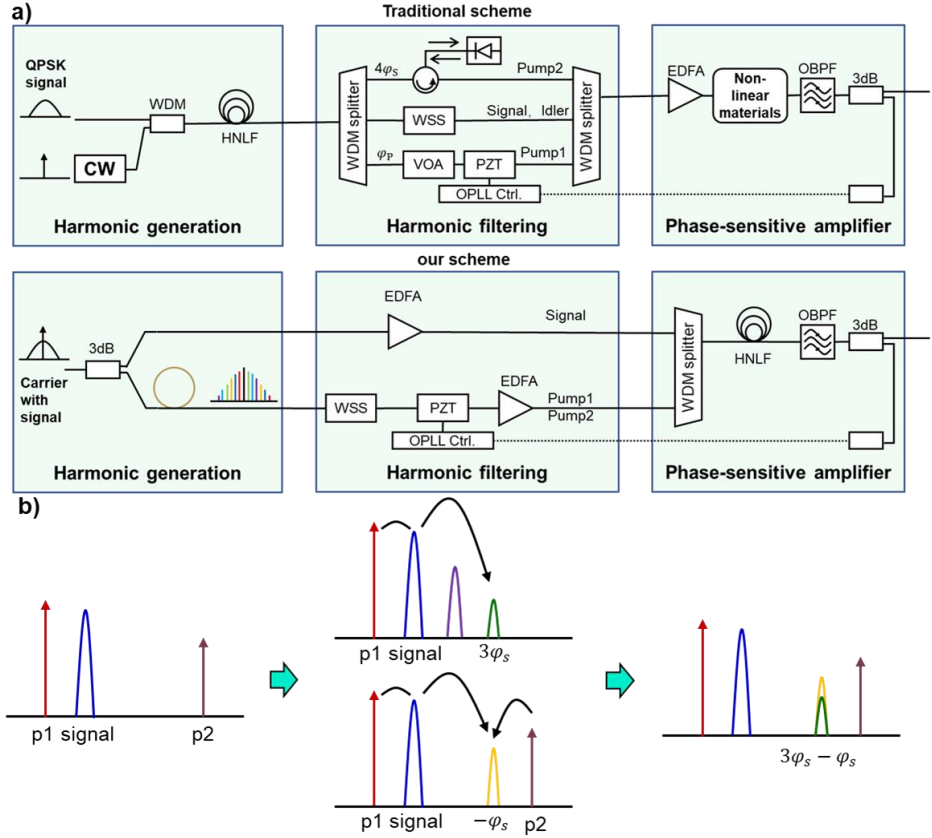

**Figure 1.** a) Comparison of traditional scheme to ours; b) The principle of single-step phase-sensitive regeneration of QPSK.

Kerr soliton combs have demonstrated high-frequency coherence and stability [15], satisfying the frequency-matching requirements for phase regeneration in high-order modulation formats. Conventional schemes for generating matching pump lights, such as electrical or optical modulation of the received signal, are often complex and limited by the numbers of the comb teeth [13,16], as shown in Figure 1a). To address this challenge, we propose a novel approach a carrier recovery scheme to generate Kerr soliton combs. This approach utilizes carrier generated through carrier recovery at the receiver as the pump source for the production of Kerr soliton combs, which have dozens to hundreds of comb teeth, high-frequency coherence characteristics, and fixed and unchanged FSRs for microcavities of the same parameters [15, 17, 18]. This approach can meet the needs of future multi-channel, high-capacity, all-optical communication systems and enable multi-channel, high-speed, high-order modulation format signal phase regeneration processes.

In this article, we demonstrate a QPSK phase-sensitive regeneration scheme utilizing HNLF through Kerr soliton combs generated by carrier recovery scheme. Our experimental results indicate that the high phase coherence and frequency coherence of the Kerr soliton combs effectively facilitate phase-sensitive regeneration of a 20 Gbaud/s QPSK signal on HNLF, without the necessity for linear strain. Furthermore, we investigate the impact of Kerr soliton combs generated from different optical carrier-to-signal ratios (OCSR) of carrier recovery on phase-sensitive regeneration of an 8 Gbaud/s QPSK signal.

## 2. Materials and Methods

Figure 1b) displays the fundamental principle of single-step phase-sensitive regeneration of QPSK [19]. HNLF receives three beams of light, which include a signal and two pumps, referred to as P$_1$ and P$_2$. The signal carries a phase signal $\varphi_s$, and the frequency difference between the signal and P$_1$ and P$_2$ is established as Δf$_1$ and Δf$_2$, respectively. To accomplish the regeneration requirements, Δf$_2$ is configured to be thrice that of Δf$_1$. The regeneration signal is directly obtained between the signal and P$_2$ during the single-step regeneration process, which is located at a frequency of Δf$_1$ away from P$_2$. In order to demonstrate the principle, the details of three light beams undergoing four-wave mixing processes simultaneously in a highly nonlinear fiber (HNLF) are presented in Figure 1b). One of these processes, occurring between P$_1$ and the signal, yields multiple harmonics, including the third harmonic that carries three times the signal phase $3\varphi_s$. The other process, taking place between the signal and the two pumps, results in an idle light that carries the negative signal phase of -$\varphi_s$ at the location of the third harmonic generated in the aforementioned process. According to the principle of phase regeneration, defined in Equation 1, a regeneration effect can be obtained by ensuring that the ratio between -$\varphi_s$ idler and $3\varphi_s$ idler equals 1/3. The 4.77 dB reduction in the $3\varphi_s$ idler, as compared to -$\varphi_s$ idler, can be achieved through the adjustment of input power for the three beams. During the digital signal processing decoding process, the regenerated QPSK signal $\phi_{s'}=\varphi_s$ can be gotten.

$$A_s(\phi) \cdot \exp(i\phi_{s'}) = \exp(i\phi_s) + \frac{1}{3}\exp(-i3\phi_s), \tag{1}$$

To simplify the intricacy of the conventional system, we employ the Kerr soliton combs to achieve QPSK regeneration. Figure 1a) displays our proposed and traditional regeneration schemes, which consist of three primary components: harmonic generation, harmonic filtering, and phase-sensitive amplification. The initial two stages differentiate the two approaches. Specifically, our method utilizes Kerr soliton combs as regenerated pumps, while the conventional method typically employs the four-wave mixing (FWM) process between P$_1$ and the signal to generate two pumps that match the signal frequency. This process passes the signal through a high nonlinear material, resulting in an idle light (I$_1$) carrying three times the signal phase $3\varphi_s$, as well as another idle light beam at P$_2$ carrying four times the signal phase $4\varphi_s$. To achieve phase regeneration, P$_2$ necessitates injection locking to produce continuous light without the signal phase, which negates the impact of four times the signal phase $4\varphi_s$ on P$_2$. Simultaneous feeding of P$_2$, signal, I$_1$, and P$_1$ into the phase regeneration material enables the generation of idle light (I$_2$) carrying negative three times the signal phase -$3\varphi_s$ through the FWM process of P$_2$, I$_1$, and P$_1$ at the signal. The superimposition of I$_2$ and the signal produces a regenerative signal carrying the signal phase of $\varphi_{s'}$. A notable drawback of this scheme includes the cumbersome injection locking process and the reliance on the fourth harmonic to generate P$_2$. It requires a sufficiently large conversion efficiency of the high nonlinear material in the first step and necessitates high P$_1$ and P$_2$ power in the regeneration process to produce I$_2$ that satisfies the regeneration requirements. Additionally, using -$3\varphi_s$ idler generated by FWM at the signal in the regeneration process demands a power difference of 4.77 dB between -$3\varphi_s$ idler(I$_2$) and the signal, which is problematic because $3\varphi_s$ idler(I$_1$) itself is often over 10 dB smaller than the signal due to being the third harmonic of FWM. Besides, the frequency drift between -$3\varphi_s$ idler(I$_2$) and the signal can't be effectively synchronized by dividing $3\varphi_s$ idler(I$_1$) and the signal into two optical amplifications, thus the use of high P$_1$ and P$_2$ power in the regeneration process is necessary. The Kerr soliton combs utilized in this study are devoid of the drawbacks noted above. The combs were produced using the auxiliary laser heating (ALH) method [20] in a silicon nitride chip, where the pumps were derived from carrier recovery. The

Kerr soliton combs exhibit uniform frequency spacing and high frequency coherence, and can directly generate P₁ and P₂ without carrying the signal phase. This eliminates the need for injection locking and simplifies the experimentation process by ensuring that the frequency interval requirements between P₁, P₂ and the signal are met. The principle of single-step QPSK signal regeneration can be utilized to obtain the regenerated signal at the third harmonic frequency produced by P₁ and the signal. This regeneration signal is generated through the coherent addition of two FWM processes, and even at a low input pump power, the regenerative effect can occur when the two idlers ($3\varphi_s$ and $\varphi_s$) satisfy the power relationship. The use of a lower input pump power can prevent the SBS effect in HNLF, reduce the interference of amplitude noise in the phase regeneration process, and facilitate phase-sensitive regeneration.

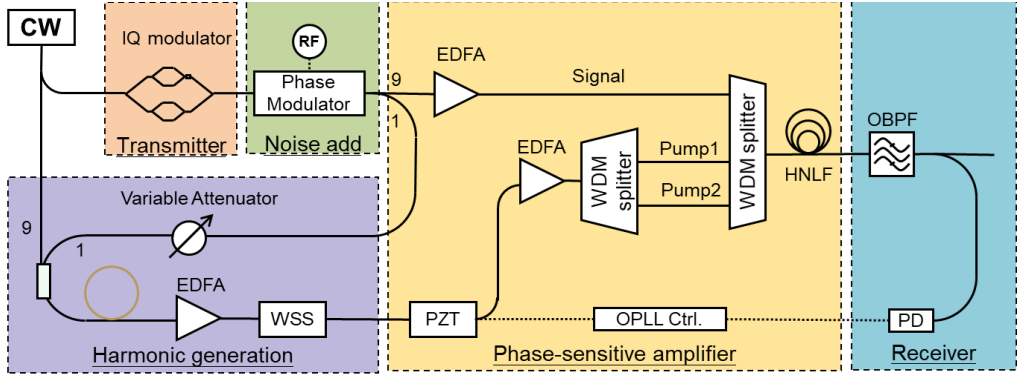

**Figure 2.** The experimental setup of single-step phase-sensitive regeneration of QPSK.

The phase regeneration of the QPSK signal was successfully achieved based on the previously mentioned scheme, as illustrated in Figure 2. A continuous wave (CW) laser with a wavelength of 1550 nm was split into two channels through a 3dB splitter. The first path was modulated by 8 or 20 GBaud/s QPSK data using an IQ modulator, whereas the second path was directed to the harmonic generation section. The generated QPSK signal was further modulated by a phase modulator driven by an electrical noise signal which is derived by an arbitrary waveform generator (AWG). Using a 1:9 optical splitter, the degraded QPSK signal was split into two channels, with the 90% portion passing through the phase-sensitive amplifier part as the signal to regenerated, while the remaining 10% was directed towards the harmonic generation section. The degraded signal that entered the harmonic generation section was merged with the continuous light, which was then split into a 1:9 optical splitter as the carrier to recover pump used for Kerr comb generation, following attenuation by a variable light attenuator. The OCSR of the carrier recovery was adjustable using the variable light attenuator. After amplification by the low-noise erbium-doped fiber amplifier (EDFA), the two comb teeth at 1549.2 nm and 1552.4 nm were filtered out using a wavelength selective switch (WSS) as P₁ and P₂, respectively. The signal attenuated differently between P₁ and P₂ to meet the power requirements of subsequent phase-sensitive amplification. Subsequently, the two light beams were transmitted to the phase-sensitive amplifier part.

In the phase-sensitive amplifier module, the dual-pump input is initially amplified by a single EDFA. Next, the amplified dual-pump are split and recombined with the amplified degraded signal using wavelength-division multiplexer (WDM) splitters and combines. This step eliminates the out-of-band noise arising from amplified spontaneous emission (ASE) noise. Finally, the signal is coupled into the highly nonlinear fiber (HNLF) with an input power of 20 dBm, while the dual-pump with an input power of 14 dBm. After regeneration, the regeneration signal at 1551.6 nm is filtered out by an optical bandpass filter, and the regeneration signal is split into two parts by a 3dB splitter. One path is converted to an electrical signal using photodetectors (PD) and then fed back to the dual pump through an electromagnetic amplification and locking circuit through the piezoelectric ceramic. The feedback loop eliminates low-frequency phase jitter caused by environmental noise, allowing for phase match of the signal and pumps, resulting in continuous regeneration. The other path of regeneration signal has been detected using an optically coherent receiver followed by oscilloscope. And The signal constellation diagram and the signal-to-noise ratio (SNR) have been measured via digital signal processing（DSP）to accurately quantify the phase noise.

## 3. Results

In order to provide a clearer demonstration of the phase-sensitive effect, Figure 3(c, d) displays the input and output spectra of the HNLF. Our experimental results demonstrate that in the unlocked state, the regeneration signal experiences a phase jitter of 3 dB at the idle frequency, caused by the random changes in the phase difference between the signal and the dual-pump due to ambient noise. These observations validate the occurrence of the phase-sensitive amplification process. The optical spectrum of the Kerr soliton comb can be observed in Figure 3(b). We adopt one $Si_3N_4$ micro-ring cavity for Kerr soliton comb generation, the cross-section of the micro-cavity is 1650×800 $nm^2$ and the FSR is around 100 GHz. The $Si_3N_4$ chip is packaged with polarization-maintaining I/O fibers, which is shown in Figure 3a).

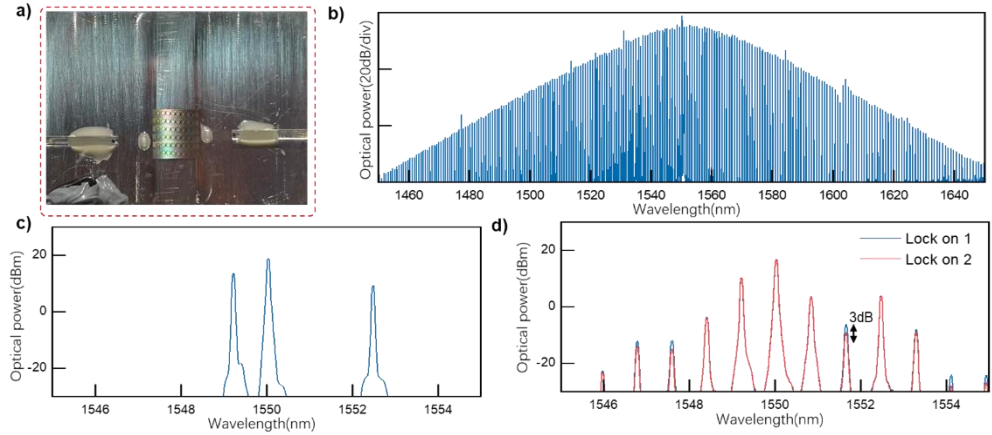

**Figure 3.** a) The pictures of packaged Si3N4 chips used in our experiment; (b) Measured optical spectrum for kerr soliton comb; c) The input spectrum of the HNLF; d) Output spectrum under conditions that satisfy different phase matching between the signal and the pump.

Our experimental results indicate that the regeneration effect is highly reliant on the frequency match between the dual-pump and the signal. The regeneration process takes place through the superimposition of $3\varphi_s$ idler and $-\varphi_s$ idler. Hence, the closer the frequencies of these two light beams, the greater the regeneration effect. Our study found that variations of the OCSR in carrier recovery had only a minor effect on the frequency stability of the comb but did affect the phase regeneration effect. Figure 4 illustrates the constellation diagrams of the 8 GBaud/s QPSK signal at different OCSR in the carrier recovery, with varying levels of phase noise. Our experiment showed a significant QPSK phase regeneration effect, leading us to conclude that the OCSR in carrier recovery has a substantial impact on the regeneration effect.

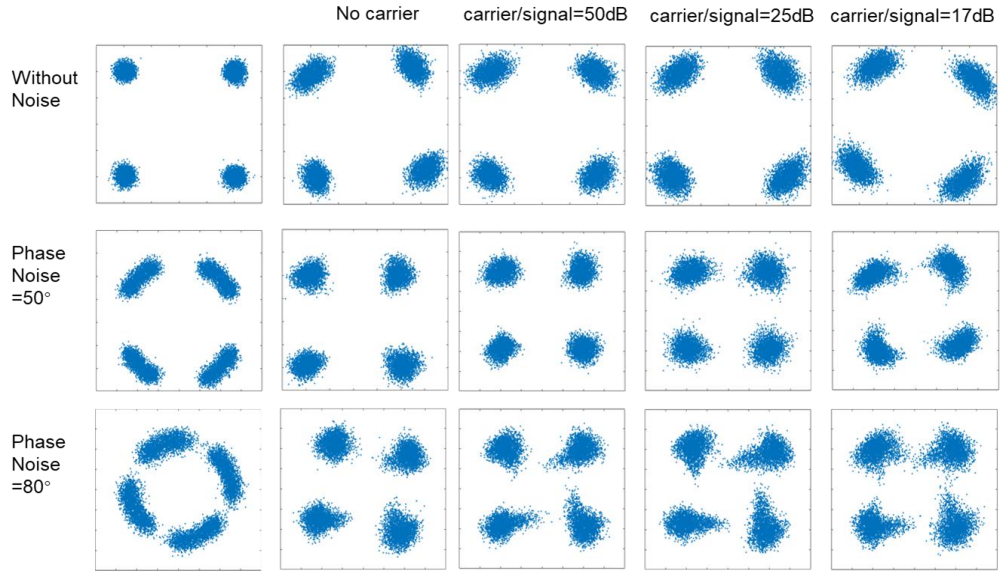

**Figure 4.** Constellation diagrams of different input carrier ratios with no noise, 50 degrees of phase noise, and 80 degrees of phase noise, respectively.

In the case of signal without degradation, the process of phase regeneration introduces amplitude noise due to its underlying mechanism. Highly compressed phase effects lead to broadening of amplitude noise. Compression of phase noise during phase regeneration is significant when the initial phase noise is between 50 and 80 degrees. However, the phase regeneration process has weaker suppression ability for amplitude noise as the initial phase noise increases. A comparison of the regeneration effect for different OCSR in carrier recovery shows that more signal carried in carrier recovery results in weaker compression of phase noise and stronger effects of amplitude noise. This is because a larger OCSR in carrier recovery causes greater disturbances to the frequency stability of the comb, which subsequently affects the coherent addition of the two idlers and impacts the regeneration effect. Nonetheless, such disturbances do not significantly hinder the effect of phase regeneration. Phase regeneration effect can still be observed even when the OCSR is 17 dB. Moreover, when the carrier is 25 dB bigger than the signal in carrier recovery, the perturbation caused by the carrier recovery on the comb frequency stability has minimal effect on the phase noise compression capacity.

In this study, we analyzed the changes in SNR of a 20 GBaud/s QPSK signal with consistent receiver input power under two different levels of phase noise, without signal carried in carrier recovery, to better demonstrate the difference in signal quality before and after regeneration. Our experimental results, as illustrated in Figure 5b), reveal an improvement in the sensitivity of the regenerated signal reception by more than 9 dB when compared to the signal before regeneration. Moreover, we conducted an intentional experiment on the impact of phase locking, and Figure 5a) displays the significant response of the locked signal to the perturbations introduced to the link.

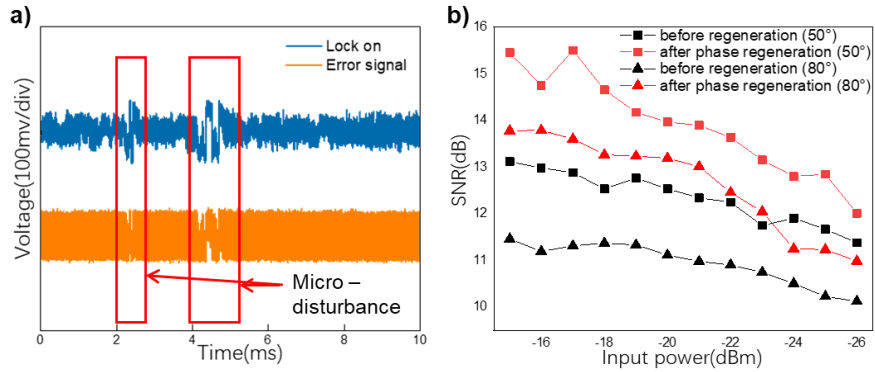

**Figure 5.** a) Lock signal (blue) and error signal (yellow) shows the phase locking effect; b) shows the calculated SNR before (black)and after regeneration(red)   .

## 4. Conclusion

This paper presents a carrier recovery scheme to generate Kerr soliton combs for phase regeneration of 20 GBaud/s QPSK signals in HNLF using a low input dual-pump power of 14 dBm. The approach increases the sensitivity of the received regenerated signal by more than 9 dB at the same receiving SNR. Furthermore, the paper compares the regeneration effects of 8 GBaud/s QPSK signals at 50° and 80° under different OCSR in carrier recovery. This comparison highlights the frequency stability requirements between the dual pumps necessary for the phase regeneration process. The study simplifies the existing experimental architecture of phase regeneration based on HNLF while demonstrating exceptional regeneration performance. Additionally, this work addresses the recovery problem of the Kerr comb in practical communication processes and proposes a regeneration scheme that can be used for quality recovery in long-distance all-optical communication in the future.

**Supplementary Materials:** The data presented in this study is available from the corresponding author upon reasonable request

**Author Contributions:** Conceptualization, Y.G. and X. H.; methodology, Y.G. and X. H.; software, X. H.; validation, Y.G., X. H. and H.K.; formal analysis, X. H.; investigation, K.Q.; resources, K.Q.; data curation, X. H.; writing—original draft preparation, X. H.; writing—review and editing, Y.G.; visualization, Y.G.; supervision, K.Q.; project administration, K.Q.; funding acquisition, K.Q. All authors have read and agreed to the published version of the manuscript.

**Funding:** This work is supported by the National Key Research and Development Program of China (2019YFB2203103, 2021YFB2800602); State Key Laboratory of Advanced Optical Communication Systems and Networks (2021GZKF010); National Natural Science Foundation of China (61705033, 62001086); Sichuan Province Science and Technology Support Program (2021YJ0095); Fundamental Research Funds for the Central Universities (2021J003); Sichuan Science and Technology Program（2022YFSY0062）.